\documentclass[%
 reprint, 
 superscriptaddress,
 bibnotes,
 amsmath,amssymb,
 aps,
 prd,
]{revtex4-2}

\usepackage[dvipdfmx]{graphicx}
\usepackage{hyperref} 
\usepackage{dcolumn}
\usepackage{bm}
\usepackage{siunitx}
\DeclareSIUnit[number-unit-product = ]\percent{\%} 
\usepackage[caption=false]{subfig} 
\usepackage[version=4]{mhchem} 

\begin{document}

\title{Measurement of Neutron Whispering Gallery States Using a Pulsed Neutron Beam}

\author{G.~Ichikawa}
\email{Corresponding author: go.ichikawa@kek.jp}
\affiliation{High Energy Accelerator Research Organization, Tsukuba 305-0801, Japan}
\affiliation{J-PARC Center, Tokai 319-1195, Japan}

\author{K.~Mishima}
\affiliation{Kobayashi-Maskawa Institute, Nagoya University, Nagoya 464-8602, Japan}
\affiliation{High Energy Accelerator Research Organization, Tsukuba 305-0801, Japan}

\date{\today}

\begin{abstract}
A neutron whispering gallery state is a quantum state localized on a material surface bound by the centrifugal force and the material potential. Precise measurements of such quantum states enable tests of quantum mechanics in non-inertial frames, characterization of the surface potential, and searches for hypothetical short-range interactions at the nanometer scale. We observed a neutron whispering gallery state on a \ce{SiO2} concave mirror using a pulsed cold neutron beam. The measured results agree with theoretical calculations within \qty{1.9}{\percent} for the centrifugal acceleration $a \approx \qty{7e7}{m/s^2}$, which is due to unmodeled deviations of the shape of the concave mirror edge from an ideal one. We found that the sensitivity itself was \num{1e-4}, which is two orders of magnitude better than the above agreement.
\end{abstract}

\maketitle

\section{Introduction\label{sec:introduction}}
General relativity, which describes gravitational interactions, and the Standard Model, which is based on quantum mechanics, explain most physical phenomena and form the foundation of modern physics. However, the unification of these two theories remains an open challenge. Experimental investigations that may provide insights into unification are indispensable. Because general relativity and quantum mechanics describe distinct regimes, only a few experiments test simultaneously. Hence, for several decades, unique physical systems where both gravity and quantum mechanics appear, realized by neutrons, have attracted considerable interest~\cite{Colella_1975,Nesvizhevsky_2002}.

The Colella-Overhauser-Werner (COW) experiments~\cite{Colella_1975} measured the gravitational phase shift in interference fringes in a Si single-crystal neutron interferometer. In this setup, the entire interferometer was rotated so that the Earth's gravitational potential introduced a phase difference between the two paths.
The initial experiment confirmed that the experimental result roughly agreed with the theoretical prediction, with a discrepancy of about \qty{12}{\percent}. The primary cause of the discrepancy was the deformation of the interferometer itself when tilted. A subsequent COW-type experiment using two nearly harmonic neutron wavelengths to correct for this deformation improved the agreement with theory to \qty{0.8}{\percent}~\cite{Littrel_1997}. Furthermore, an experiment in which the interferometer was subjected to accelerated motion successfully tested the equivalence of gravity and acceleration with \qty{4}{\percent} precision, contributing to the verification of the equivalence principle at the quantum level~\cite{Bonse_1983}, one of the fundamental principles of general relativity.

Gravitationally bound neutron states offer an alternative approach to studying quantum effects in gravity. When the vertical energy of a neutron is below the Fermi potential of a material ($\approx \qty{100}{neV}$), total reflection from a surface forms a state localized to approximately \qty{10}{\micro\meter} above the mirror.
The gravitationally bound states were first observed using ultracold neutrons~\cite{Nesvizhevsky_2002}. In that experiment, a ceiling designed to remove neutrons was adjusted to different heights while measuring the neutron count. It was confirmed that neutrons survived only when the ceiling height exceeded approximately \qty{15}{\micro\meter}, corresponding to the extent of the ground-state wave function, which demonstrated the existence of the ground state bound by gravity.
Subsequently, the qBOUNCE experiment induced Rabi transitions among these discrete levels with a vibrating floor~\cite{Jenke_2011}.
Neutrons that transition to higher energy levels are selectively removed by a ceiling, resulting in a marked reduction in the neutron counts at specific transition frequencies. This method allows for precise measurement of the energy-level splitting. It has been demonstrated to achieve a relative precision of \num{4e-3} against Earth's gravity~\cite{Cronenberg_2018}.

These experimental results play an important role in testing gravity and quantum mechanics and, notably, impose unique constraints on theories of gravity beyond general relativity, including noncommutative structures of spacetime~\cite{Saha_2014}, the standard-model extension~\cite{Escobar_2022}, beyond-Riemann gravitational theories~\cite{Kostelecky_2021, Ivanov_2021}, and entropic gravity~\cite{Verlinde_2011, Chaichian_2011, Schimmoller_2021}.
Moreover, because the wave function of gravitationally bound neutrons is localized near the mirror, such experiments can search for hypothetical short-range interactions~\cite{Murata_2015, Sponar_2021, Antoniadis_2011}.
The qBOUNCE experiment~\cite{Jenke_2014,Cronenberg_2018} also searched for interactions predicted by axions~\cite{Moody_1984} and screened dark energy~\cite{Khoury_2004,Hinterbichler_2010}.

\begin{figure*}[!tbhp]
\includegraphics[keepaspectratio,width=0.98\linewidth]{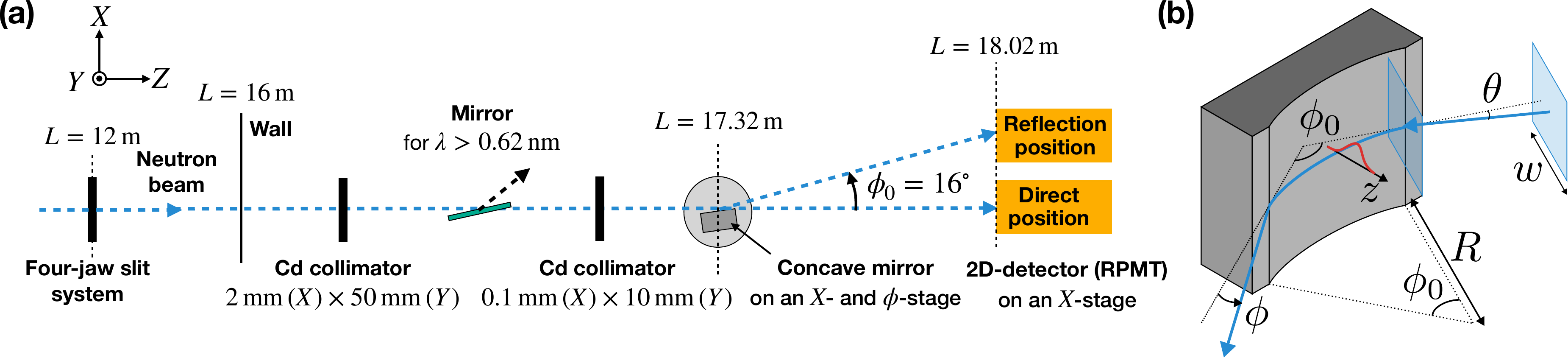}
\caption{(a) Schematic top view of the experimental geometry, where $L$ denotes the distance from the neutron moderator.
(b) Enlarged view of the concave mirror and the neutron whispering gallery state (red curve). The dimensions and shapes are not drawn to scale. The blue rectangle represents the neutron beam cross section, and the arrow indicates a neutron trajectory. The concave surface has a radius of curvature $R$ and an angular span $\phi_0$. A neutron beam with a width of $w$ enters the mirror at an incidence angle $\theta$, and a neutron becomes bound as a neutron whispering gallery state on the mirror, where $z$ denotes the height from the mirror surface, exiting at a deviation angle $\phi$.
}
\label{fig:geometry_concave}
\end{figure*}

Due to the equivalence of gravity and acceleration, quantum states bound by centrifugal force can be regarded as analogs of gravitationally bound states. The quantum states bound by the centrifugal force and a mirror are called neutron whispering gallery states~\cite{Nesvizhevsky_2010}. In an experiment with cold neutrons from a research reactor, interference fringes produced by these states localized to around \qty{100}{nm} near the mirror surface were observed.
Because neutron whispering gallery states undergo hundreds of reflections from the mirror surface in the classical picture, they are highly sensitive to the shape of the potential.
Therefore, measuring these states can test the surface potential shape and enable searches for hypothetical interactions at distances even closer than those accessible by gravitationally bound states.
Additionally, by comparing with measurements of gravitationally bound states, one can test the equivalence principle in quantum mechanics~\cite{Zych_2018} and investigate quantum mechanics in non-inertial frames with large accelerations~\cite{Nauenberg_2016}.

This paper reports a precise measurement of neutron whispering gallery states using a pulsed neutron beam for the first time. The paper is organized as follows. Section~\ref{sec:experiment} explains the experimental setup and measurement methods. Section~\ref{sec:concave} describes how the radius of curvature and angular span of a concave mirror made of \ce{SiO2} were determined based on surface measurements. In Section~\ref{sec:formalism}, we derive the theoretical formalism for neutron whispering gallery states on the concave mirror and show how the observed distributions are calculated. Section~\ref{sec:analysis} presents the analysis of the experimental data, comparison with theoretical calculations, and discussions. Finally, Section~\ref{sec:conclusion} provides conclusions and outlook.

\section{Experiment\label{sec:experiment}}
This experiment was conducted at the low divergence branch of the BL05 beamline~\cite{Mishima_2009,Mishima_2015} of the Materials and Life Science Experimental Facility (MLF)~\cite{Nakajima_2017} at the Japan Proton Accelerator Research Complex (J-PARC).
At J-PARC, a \qty{3}{GeV} proton beam is directed onto a mercury target at a repetition rate of \qty{25}{Hz}, generating neutrons through nuclear spallation reactions. During this experiment, the beam power was maintained at approximately \qty{700}{kW}. Cold neutrons, which have been slowed down by a moderator, are then guided to the beam line.
The use of a pulsed neutron source, such as J-PARC, enables the determination of the neutron wavelength distribution via the time-of-flight method without reducing the available statistics, unlike when using a chopper.

A schematic top view of the apparatus used in this experiment is shown in Fig.~\ref{fig:geometry_concave}(a).
In the following description, the distance from the neutron moderator is denoted by $L$ and the components are explained from upstream to downstream along the beamline.
The cold neutron beam delivered to the beamline is first shaped by a four-jaw slit system located at $L=\qty{12}{m}$. In this experiment, the horizontal beam width (the $X$ direction in the figure) was fixed at \qty{6}{mm}, providing a horizontal divergence with a standard deviation of \qty{0.09}{mrad}. The vertical beam width (the $Y$ direction in the figure) was set to \qty{44}{mm}, which results in vertical divergence of \qty{5}{mrad}.

The beam travels through a vacuum guide tube up to $L=\qty{16}{m}$. A Cd collimator with an opening of $\qty{2}{mm}\,(X)\times\qty{50}{mm}\,(Y)$ was placed at $L=\qty{16.49}{m}$, and a neutron mirror that reflects and removes neutrons with wavelength $\lambda>\qty{0.62}{nm}$ was positioned at $L=\qty{17.04}{m}$ to prevent detection of delayed neutrons from the previous pulse. A second Cd collimator with an opening of $\qty{0.1}{mm}\,(X)\times\qty{10}{mm}\,(Y)$ was placed at $L=\qty{17.17}{m}$. This final collimator defined the beam size entering the concave mirror. A glass concave mirror mounted on an $X$- and $\phi$-stages ($\phi$ denotes yaw-rotation) was placed at $L=\qty{17.32}{m}$. Floor vibrations were suppressed by an active vibration isolation table. The concave mirror was made of OHARA QUARTZ SK-1300B (\ce{SiO2}, density \qty{2.2}{g/cm^3}, surface potential $U_0=\qty{90.5}{neV}$), and its concave surface was fabricated via magneto-rheological finishing by Crystal Optics Inc. with high precision.
A \qtyproduct{15x25x40}{mm} rectangular piece of glass had one of its \qtyproduct{25x40}{mm} faces formed into a concavity about the long edge as an axis.
We selected \ce{SiO2} to avoid the diffuse boundary created by the oxide layer on the surface, as reported for a Si mirror~\cite{Nesvizhevsky_2010_NJP}, and also because \ce{SiO2} is well-established for precise polishing.
The ideal conditions for a concave mirror require that the surface roughness be significantly smaller than the length scale of the whispering gallery state (\qty{30}{nm} at $\lambda=\qty{0.3}{nm}$) and that the surface waviness be much smaller than the angle determined by the ratio of the velocity scale to the tangential velocity (\qty{1.6}{mrad} at $\lambda=\qty{0.3}{nm}$).

An expanded view of the concave mirror and the relevant parameters used in this experiment are shown in Fig.~\ref{fig:geometry_concave}(b). A portion of the incoming neutron beam is bound by the material potential and the centrifugal potential, forming a whispering gallery state that travels along the mirror surface. The beam direction is bent by the angular span $\phi_0$ of the concave mirror. Other parameters are: $R$, the radius of curvature of the concave mirror; $w$, the beam width; $\theta$, the incidence angle of the beam relative to the mirror's tangential direction; $z$, the height from the mirror surface toward the center of curvature; and $\phi$, the deviation angle of the outgoing neutrons.
The concave mirror was designed with a radius of curvature $R=\qty{25}{mm}$ and an angular span $\phi_0=\qty{16}{\degree}$, selected to optimize the observation of the ground state and interference fringes of the lowest two energy levels with the wavelength distribution of the BL05 beamline.

The radius of curvature and angular span were measured by a three-dimensional profilometer (UA3P, Panasonic). Details of the measurement procedure are presented in Section~\ref{sec:concave}. The average surface roughness ($R_a$) was measured by a three-dimensional optical surface profiler (NewView, Zygo) over an area of \qtyproduct{0.08x0.08}{mm}, giving $R_a=\qty{0.58}{nm}$.
The measured surface roughness is sufficiently smaller than the characteristic length scale of the neutron whispering gallery state (\qty{30}{nm}).

A two-dimensional detector with time resolution, utilizing a resistance-division photomultiplier tube (RPMT)~\cite{Hirota_2005}, on an $X$-stage was placed at $L=\qty{18.02}{m}$.
The neutron whispering gallery state was observed as the distribution of $\lambda$ and $\phi$ on the RPMT, measured at a position where the RPMT was shifted \qty{199}{mm} in the $+X$ direction.
We defined a foreground region of $\qty{30}{mm}\,(X)\times\qty{21}{mm}\,(Y)$ on the RPMT plane and background regions of $\qty{30}{mm}\,(X)\times\qty{10.5}{mm}\,(Y)$ above and below the foreground region for the neutron whispering gallery measurement.

During neutron whispering gallery measurements, after ensuring that the neutron beam was incident on the entrance of the concave surface by adjusting the $X$- and $\phi$-stages, the rotation angle of the concave mirror was scanned in \qty{0.03}{\degree} increments at five points, each measured for 5 hours, to find the angle that maximized neutron counts. The angle with the highest neutron count was used for the analysis of the neutron whispering gallery state. To compare with the theoretical curve, the wavelength distribution of neutrons directly incident on the detector without the concave mirror was measured and used as a normalization. In that reference measurement, the vertical beam width was reduced to \qty{4}{mm} to avoid detector pileup. To reproduce an effective \qty{44}{mm} opening, we shifted the \qty{4}{mm} beam eleven times across the full span, and all data were summed to obtain the final distribution.

\section{Geometry of Concave Mirror\label{sec:concave}}
\begin{figure*}[tbhp]
\centering
\includegraphics[keepaspectratio,width=0.98\linewidth]{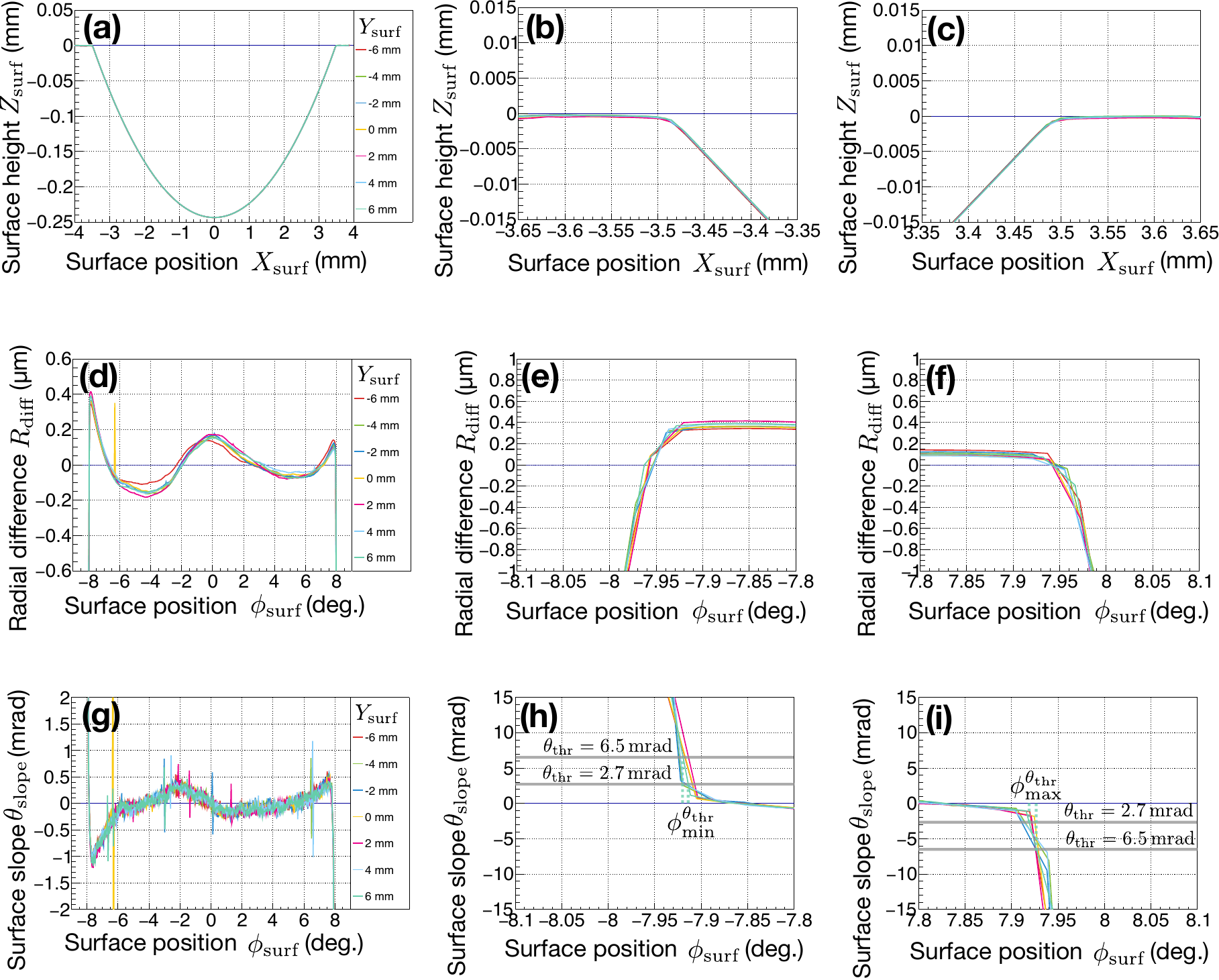}
\caption{Surface metrology data of the concave mirror. (a)--(c) show the three-dimensional surface profiles; (d)--(f) show the deviations from the circular fits; (g)--(i) show the slope angles derived from these deviations. (a), (d), and (g) cover the entire concave surface, while the others provide magnified views of the edges. The angle values $\phi_{\mathrm{min}}^{\theta_{\mathrm{thr}}}$ and $\phi_{\mathrm{max}}^{\theta_{\mathrm{thr}}}$ are used to determine the mirror's angular span; (h) and (i) show an example at $Y_{\mathrm{surf}}=\qty{6}{mm}$.}
\label{fig:surf}
\end{figure*}

This section describes how to determine the radius of curvature $R$ and the angular span $\phi_0^{\mathrm{surf}}$ of the concave mirror from surface measurements. Figures~\ref{fig:surf}(a)--(c) show the three-dimensional profile measurement of the mirror surface. We defined the measurement plane as $X_{\mathrm{surf}}Y_{\mathrm{surf}}$ and measured the height $Z_{\mathrm{surf}}$ of the glass surface to obtain the concave shape. The surface was scanned along the $X_{\mathrm{surf}}$ axis every \qty{2}{mm} in the $Y_{\mathrm{surf}}$ direction. Here, $Y_{\mathrm{surf}}=0$ corresponds to the center of the region hit by the neutron beam. In this experiment, we aligned the $Y$ axis of Fig.~\ref{fig:geometry_concave}(a) with the $Y_{\mathrm{surf}}$ axis of the surface measurement, so that the neutron beam entered from the $+X_{\mathrm{surf}}$ side.

Figures~\ref{fig:surf}(d)--(f) plot the deviations of the surface measurement data from a circular fit for each scan line. The deviations are defined as negative if the surface lies outside (farther from the center) than the fitted circle. The $X_{\mathrm{surf}}$ coordinates are converted to $\phi_{\mathrm{surf}}$, the angular coordinate on the mirror surface. We fit the obtained radii $R$ with a linear function of $Y_{\mathrm{surf}}$ and evaluate the change in $R$ between $Y_{\mathrm{surf}}=\qty{-6}{mm}$ and \qty{6}{mm}, which corresponds to the region illuminated by the neutron beam, as an uncertainty. This yields $R=\qty{25.168(8)}{mm}$.

Figures~\ref{fig:surf}(g)--(i) show the slope of the surface profile, obtained by converting deviations from the ideal circular fit into slope angles.
This distribution corresponds to the surface waviness and does not satisfy the ideal requirement for a concave mirror, i.e., being sufficiently smaller than \qty{1.6}{mrad}, which could explain the observed differences between the calculated and measured results in Section~\ref{sec:analysis}.
We determine the angular span of the mirror $\phi_0^{\mathrm{surf}}$ by defining the region where the absolute value of the slope does not exceed a threshold $\theta_{\mathrm{thr}}$.
The angular span is given by $\phi_0^{\mathrm{surf}} = \phi_{\mathrm{max}}^{\theta_{\mathrm{thr}}} - \phi_{\mathrm{min}}^{\theta_{\mathrm{thr}}}$, where $\theta_{\mathrm{thr}}$ is the positive threshold angle, and $\phi_{\mathrm{min}}^{\theta_\mathrm{thr}}$ ($\phi_{\mathrm{max}}^{\theta_\mathrm{thr}}$) is the surface position in the vicinity of the mirror edge in Figs.~\ref{fig:surf}(h) or (i) where the slope graph crosses $\theta_\mathrm{thr}$ ($-\theta_\mathrm{thr}$). In Figs.~\ref{fig:surf}(h) and (i), the data at $Y_{\mathrm{surf}}=\qty{6}{mm}$ are shown as an example. The threshold angle $\theta_{\mathrm{thr}}$ was determined according to the deviation angle $\phi$ presented in Section~\ref{sec:analysis}.
Specifically, the neutron whispering gallery distribution first appears at $\lambda=\qty{0.26}{nm}$, and neutrons with wavelengths up to $\lambda = \qty{0.62}{nm}$ where neutrons pass through the upstream mirror unfiltered. We took the corresponding classical cutoff angles of \qty{2.7}{mrad} (for \qty{0.26}{nm}) and \qty{6.5}{mrad} (for \qty{0.62}{nm}) as the thresholds $\theta_{\mathrm{thr}}$. Using these thresholds, the averages and standard deviations of the angular span were obtained as \qty{15.850(18)}{\degree} and \qty{15.8603(18)}{\degree}, respectively. We adopted their average as $\phi_0^{\mathrm{surf}}=\qty{15.86(2)}{\degree}$.

\section{Formalism\label{sec:formalism}}
In this section, we formulate the neutron whispering gallery state based on quantum mechanics~\cite{Nesvizhevsky_2010_NJP} and derive the ideal distribution observed in our experiment. In cylindrical coordinates near the mirror surface, ignoring the axial parameter due to the trivial symmetry, the one-dimensional Schr\"odinger equation can be written as
\begin{equation}
    \left[ -\frac{\hbar^2}{2m}\frac{\mathrm{d}^2}{\mathrm{d} z^2} + U_0 \Theta(-z) + \frac{mv^2 z}{R} - E_n \right]\psi_n(z)=0,
    \label{eq:Schroedinger}
\end{equation}
where $\hbar$ is the reduced Planck constant, $m$ is the neutron mass, $z$ is the displacement from the mirror surface toward the center of curvature, $U_0=\qty{90.5}{neV}$ is the optical potential of \ce{SiO2}, $\Theta$ is the Heaviside step function indicating the region of the mirror, $R$ is the radius of curvature, $v$ is the neutron velocity, $n$ is the principal quantum number, $E_n$ is the radial energy, and $\psi_n$ is the wave function. The third term on the left side corresponds to the centrifugal potential $ma z$ due to the centrifugal acceleration $a=v^2/R$. Although $\psi_n$ or $E_n$ depends on $v$ (or $\lambda$), here we omit explicit notation of $v$.

\begin{figure}[tbhp]
\centering
\includegraphics[keepaspectratio,width=0.85\linewidth]{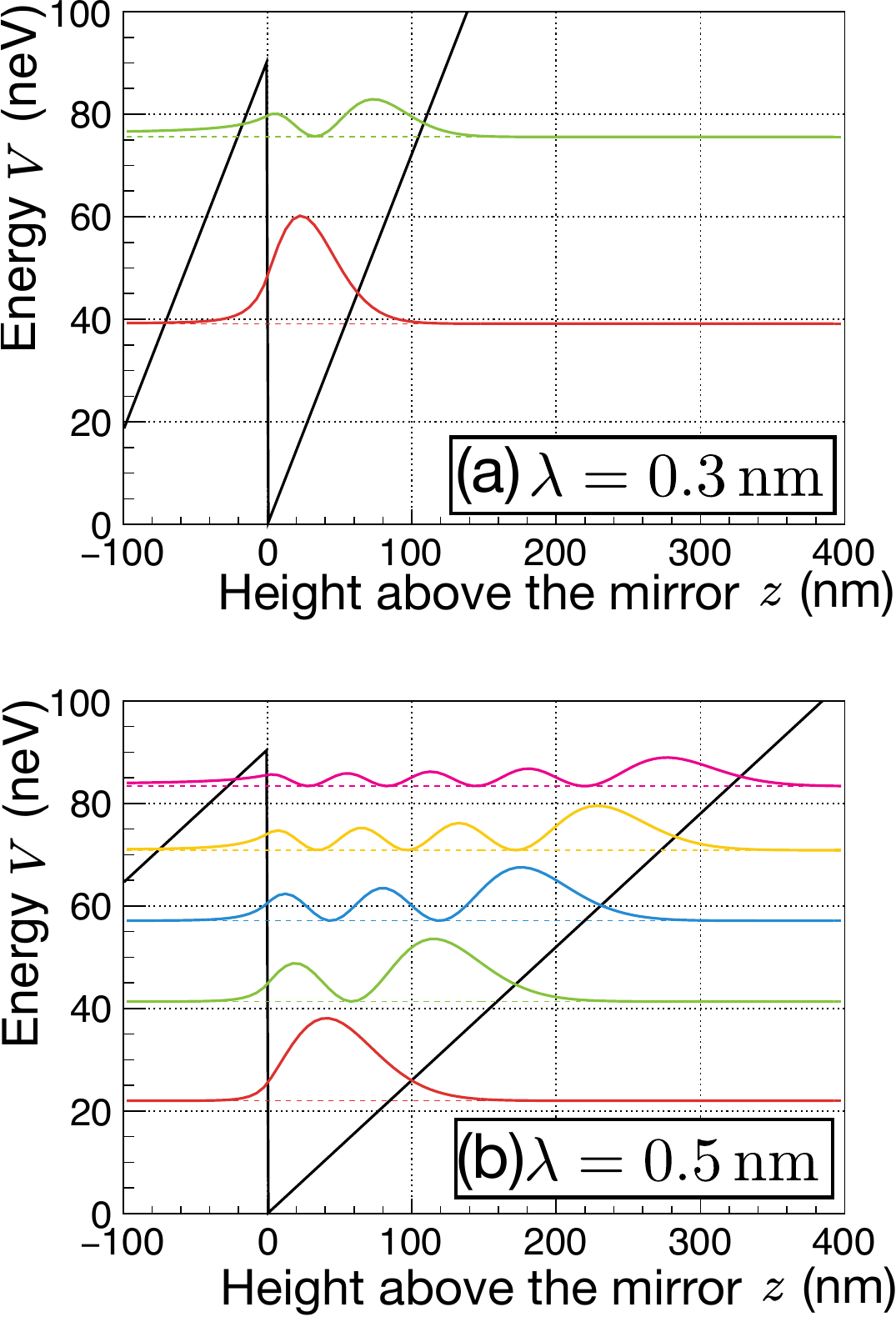}\\
\caption{Eigenfunctions and eigenenergies of neutron whispering gallery states for $\lambda=\qty{0.3}{nm}$ (a) and $\qty{0.5}{nm}$ (b). The horizontal axis is $z$, the displacement from the mirror surface toward the center of curvature, and the vertical axis is energy. The black solid line is the sum of the mirror optical potential and the centrifugal potential, the horizontal dotted lines show the real part of each eigenenergy, and the curves represent the absolute squares of the corresponding eigenfunctions. Only bound states with real eigenenergies below the mirror potential are plotted.}
\label{fig:wavefunctions}
\end{figure}

We define the characteristic length, energy, momentum, and time scales for this system as
\begin{equation}
    z_0 = \left(\frac{\hbar^2}{2m^2a}\right)^{1/3}, \quad
    E_0 = maz_0 = \left(\frac{ma^2\hbar^2}{2}\right)^{1/3},
    \label{eq:units_1}
\end{equation}
\begin{equation}
    p_0 = \sqrt{2 m E_0}, \quad t_0 = \hbar/E_0.
    \label{eq:units_2}
\end{equation}
By normalizing $z$ and $E_n$ as $\zeta=z/z_0$ and $\epsilon_n=E_n/E_0$, respectively, Eq.~\eqref{eq:Schroedinger} becomes
\begin{equation}
    \left[ -\frac{\mathrm{d}^2}{\mathrm{d} \zeta^2} + u \Theta(-\zeta) + \zeta - \epsilon_n\right] \psi_n (\zeta)=0,
\end{equation}
where $u=U_0/E_0$, and $\psi_n$ is redefined for the dimensionless variable. Examples of numerically computed eigenfunctions and eigenenergies for $R=\qty{25}{mm}$ at $\lambda = \qty{0.3}{nm}$ and \qty{0.5}{nm} are shown in Fig.~\ref{fig:wavefunctions}. Each eigenstate is quasi-stable due to tunneling into the region $z<0$; thus its eigenenergy can be written as $\epsilon_n=\beta_n - \mathrm{i}\gamma_n$, where $\beta_n$ is the real part and $\gamma_n$ is the imaginary part. As shown in Fig.~\ref{fig:wavefunctions}, the slope of the centrifugal potential varies with the neutron wavelength, and the number of surviving states increases as the wavelength increases.

When a single neutron enters the mirror edge, the wave function is
\begin{equation}
    \Psi(\zeta, \tau=0) = \sum_n a_n \psi_n(\zeta),
    \label{eq:Psi_entrance}
\end{equation}
where $\tau=t/t_0$ is the dimensionless time elapsed after incidence on the mirror, and $a_n$ is the complex coefficient giving the weight and phase of each eigenstate. Immediately before incidence on the mirror edge, the neutron can be treated as a monochromatic wave in the radial direction, since the angular divergence of the beam (\qty{0.09}{mrad}) is much smaller than the divergence of the whispering gallery state (see Section~\ref{sec:analysis}). The dimensionless radial momentum can be written as $\xi_\theta = -mv\sin\theta/p_0$ using the incidence angle $\theta$.
Because only the region near the mirror surface contributes to forming the whispering gallery state and the component directly entering $\zeta<0$ is refracted and does not form a bound state, we can set the beam width region as $0\leq \zeta < l=w/z_0$. Then, the incident wave function is written as
\begin{equation}
    \psi_{\theta}(\zeta) =
    \begin{cases}
        \exp\left(\mathrm{i} \zeta \xi_\theta \right)/\sqrt{l} & \text{$0 \leq \zeta <l$} \\
        0 & \text{otherwise}.
    \end{cases}
    \label{eq:Psi_planewave}
\end{equation}
By assuming the continuity of the wave function at the mirror entrance, we have
\begin{equation}
    a_n = \int_0^l \psi_n^*(\zeta)\,\psi_{\theta}(\zeta)\,\mathrm{d}\zeta.
\end{equation}

The wave function at the mirror exit is given by the time evolution of Eq.~\eqref{eq:Psi_entrance}:
\begin{equation}
    \Psi(\zeta, \tau_0)=\sum_n c_n \psi_n(\zeta)
    \exp\left(-\mathrm{i} \beta_n \tau_0 \right),
\end{equation}
where $\tau_0=(R \phi_0 / v )/t_0$ and $c_n = a_n \exp\left( -\gamma_n \tau_0 \right)$. In this experiment, we measured the deviation angle $\phi$, so we must consider the momentum eigenfunctions $\varphi_n$ related to $\psi_n$ by Fourier transformation. The momentum-space wave function at the exit is
\begin{equation}
    \Phi(\xi, \tau_0)=\sum_n c_n \varphi_n(\xi)
    \exp\left(-\mathrm{i} \beta_n \tau_0 \right),
\end{equation}
where $\xi=m v_r/p_0$ is the dimensionless radial momentum and $v_r$ is the radial velocity.
Using the relation of the deviation angle $\phi=\arctan\left(v_r/v\right)$, the probability distribution of $\lambda$ and $\phi$ is
\begin{equation}
    \begin{split}
        P(\lambda,\phi)
        &=\left|\Phi\left(\xi_\phi,\tau_0\right)\right|^2 \\
        &=\sum_n \left|c_n\right|^2 \left|\varphi_n(\xi_\phi)\right|^2 \\
        &\quad+ \sum_{n\neq m} c_m^* c_n  \varphi_m^*(\xi_\phi) \varphi_n(\xi_\phi)
        \exp\left[-\mathrm{i}\left(\beta_n-\beta_m\right)\tau_0\right],
        \label{eq:Phi2}
    \end{split}
\end{equation}
where $\xi_\phi = 2\pi\hbar\tan\phi/(p_0 \lambda)$
and $\tau_0 = R\phi_0 m \lambda/(2\pi\hbar t_0)$.

\section{Analysis\label{sec:analysis}}
\begin{figure}[tbhp]
\centering
\includegraphics[keepaspectratio,width=0.9\linewidth]{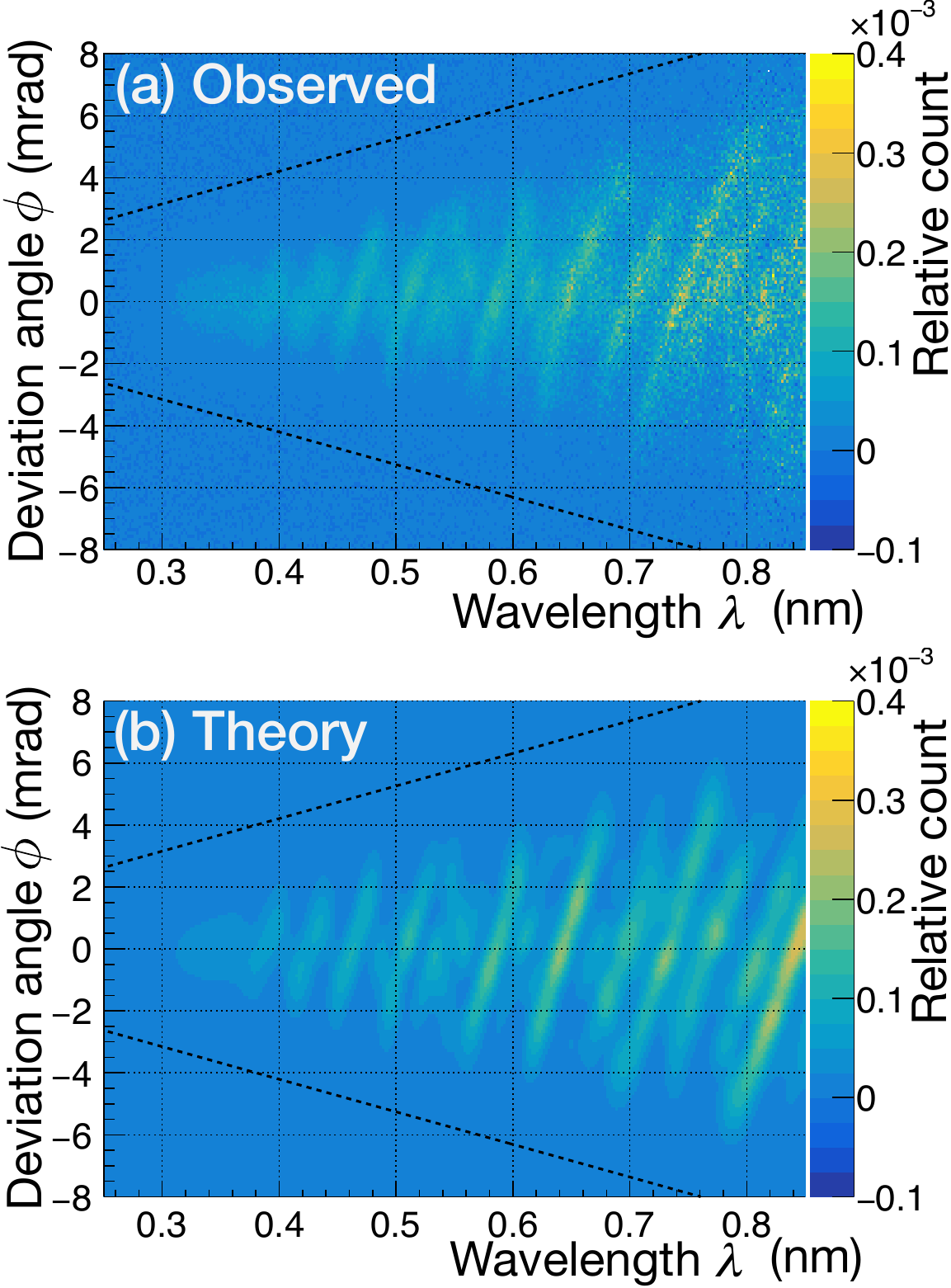}
\caption{(a) Observed and (b) theoretical two-dimensional distributions of neutron whispering gallery as functions of $\lambda$ and $\phi$. The bin sizes are \qty{0.002}{nm} and \qty{0.1}{mrad}, respectively. The diagonal dotted line corresponds to the classical cutoff angle defined by the mirror potential energy. The theory plot shown includes the additional loss term and best-fit parameters described in the text.}
\label{fig:2D}
\end{figure}

The data obtained from the neutron whispering gallery measurements were normalized by the wavelength distribution measured without the concave mirror and then compared with the theoretical distribution.
Figures~\ref{fig:2D}(a) and (b) show the $\lambda$-$\phi$ two-dimensional histograms of the observed and theoretical distributions of the neutron whispering gallery.

The wavelength $\lambda$ was obtained from the time-of-flight $T$, neutron mass $m$, and flight distance $L_F$ as $\lambda=T/(2\pi\hbar m L_F)$, with a high-precision correction measured by Bragg reflections~\cite{Shimizu_2024}.
Due to neutron moderation, the wavelength distribution measured by the time-of-flight method appears \qty{0.8}{\percent} longer than the actual wavelength. Additionally, the neutron wavelengths detected at a certain time are subject to broadening with a standard deviation of \qty{0.8}{\percent}.
Hence, the distribution $P(\lambda,\phi)$ in Eq.~\eqref{eq:Phi2} was convoluted with a Gaussian function of width $0.008\,\lambda$.

\begin{figure}[tbhp]
\centering
\includegraphics[keepaspectratio,width=0.90\linewidth]{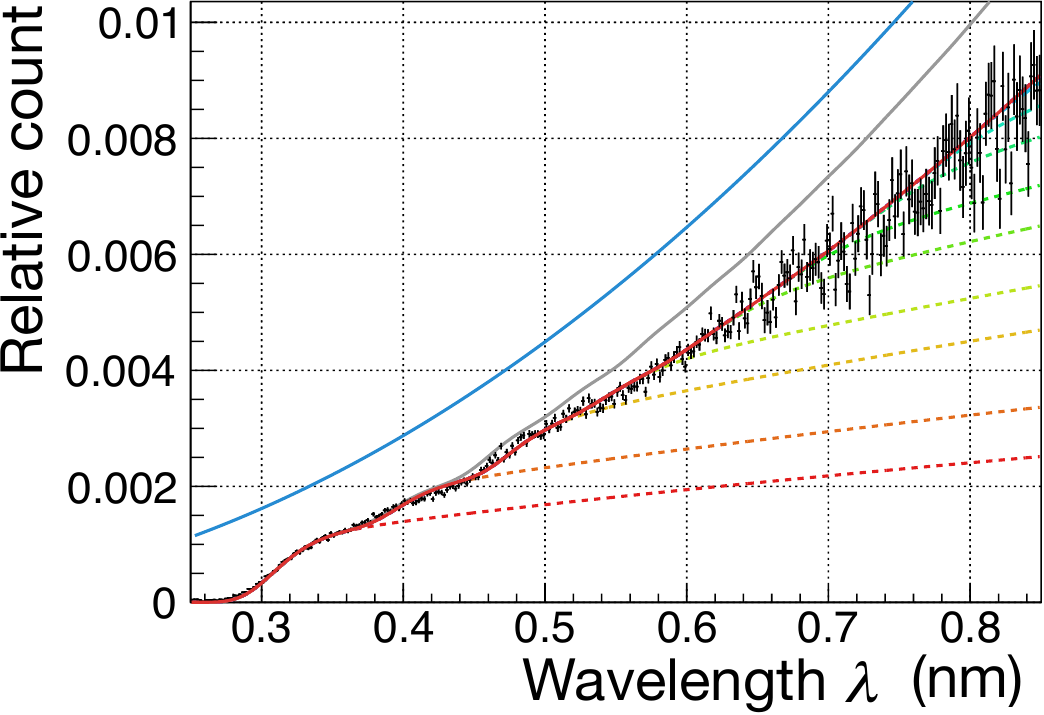}
\caption{Wavelength distribution of the neutron whispering gallery.
Black data points represent the measured data, while the blue curve corresponds to the classical theory, the gray curve is the quantum theory without additional losses, and the red curve is the quantum theory including additional losses.
The colored dashed lines illustrate the contribution of $n\le1, 2, \ldots$ states to the red curve. The theoretical curves use best-fit parameters.}
\label{fig:1D}
\end{figure}

The deviation angle $\phi$ was defined as the difference between the measured scattering angle and the mean scattering angle corresponding to $\phi_0$.
The angular span $\phi_0$ can be determined not only from surface measurements but also from the neutron scattering angle, and the value obtained by this method is defined as $\phi_0^{\mathrm{geom}}$.
We determined $\phi_0^{\mathrm{geom}}$ using the $\qty{0.30}{nm}\leq\lambda<\qty{0.35}{nm}$ region of the whispering gallery measurement which corresponds to the ground state.
Then, we obtained $\phi_0^{\mathrm{geom}}=\qty{16.06(2)}{\degree}$.
The value $\phi_0^{\mathrm{geom}}$ differs by \qty{1.3}{\percent} from the surface-measurement value of $\phi_0^{\mathrm{surf}}=\qty{15.86(2)}{\degree}$, exceeding their stated systematic uncertainty. This discrepancy likely arises from the fact that the mirror surface is slightly different from the ideal concavity near its edges, allowing the neutron wave function to follow the surface shape and scatter to a larger angle in a manner not fully accounted for in our calculation.

Figure~\ref{fig:1D} shows the wavelength distribution obtained by projecting the neutron whispering gallery data onto the $\lambda$ axis. We used $\phi_0^{\mathrm{geom}}=\qty{16.06}{\degree}$ for the theoretical calculation here. The rising edge at $\lambda=\qty{0.26}{nm}$, corresponding to the mirror potential $U_0=\qty{90.5}{neV}$, and the staircase-like increase due to the increasing number of bound states are in good agreement between experiment and quantum theory. However, the measured counts are lower than the theoretical curve at longer wavelengths, suggesting a wavelength-dependent neutron loss from the bound states.

It is known that neutron reflection at the surface potential typically leads to a loss of about $10^{-4}$ per bounce~\cite{Steyerl_1977}, which does not reproduce the data here because the loss would become larger at shorter wavelengths, opposite to the observed behavior. If the material surface contains impurities, the potential near the floor is effectively a sigmoidal ``soft potential''~\cite{Scheckenhofer_1977}, reducing the reflectivity. However, such a soft potential significantly shifts the rise at $\lambda=\qty{0.26}{nm}$ in Fig.~\ref{fig:1D} and the interference fringes in Fig.~\ref{fig:2D}, failing to match our data. A classical model where the radial energy of the neutron changes from the reflection of a tilted floor also fails to reproduce the data because the resulting energy shift changes the position of the interference fringes.

\begin{figure*}[tbhp]
\centering
\includegraphics[keepaspectratio,width=0.98\linewidth]{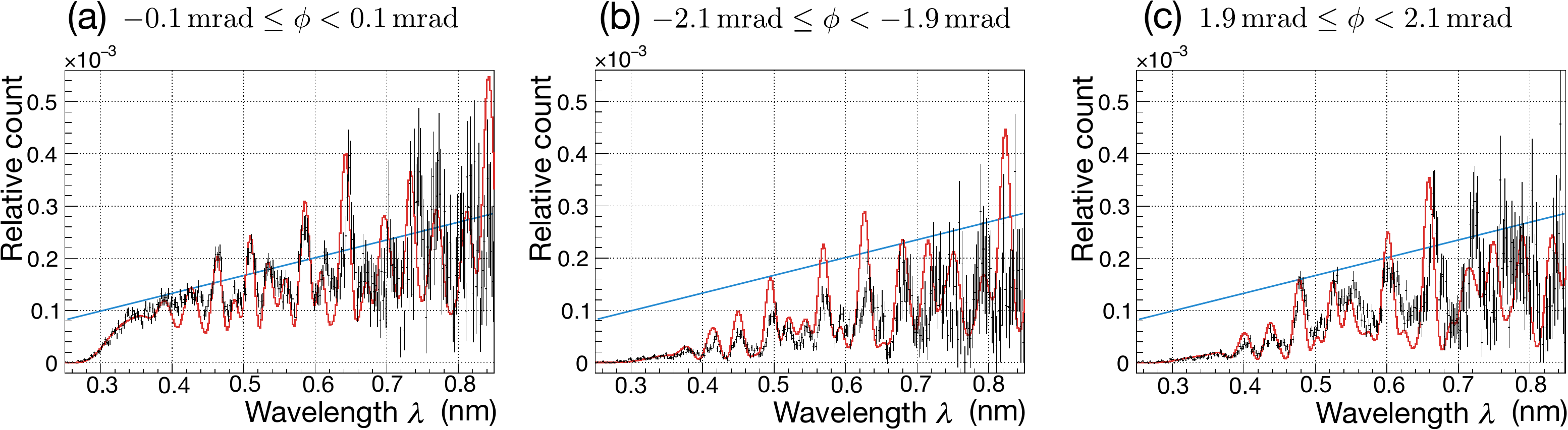}
\caption{Wavelength distributions for (a) $\qty{-0.1}{mrad}\leq\phi<\qty{0.1}{mrad}$, (b) $\qty{-2.1}{mrad}\leq\phi<\qty{-1.9}{mrad}$, and (c) $\qty{1.9}{mrad}\leq\phi<\qty{2.1}{mrad}$.
The black points represent the experimental data, the blue curves denote the classical calculation, and the red curves correspond to the quantum model including additional losses.
The interference fringes were fitted under the condition in (a), and the same best-fit parameters are used in (b) and (c).}
\label{fig:1D_slice}
\end{figure*}

The fact that more levels exist below $U_0$ in the longer wavelength region as illustrated in Fig.~\ref{fig:wavefunctions} suggests that the additional loss occurs for levels with larger $n$.
To achieve such an effect, we introduced an additional loss model in which the mirror potential height fluctuates by $\Delta$, for example, due to surface roughness or waviness, so that neutrons sometimes experience a reduced potential and escape the bound state upon reflection. In a classical picture, for the $n$-th bound state with a reflection period
\begin{equation}
    t_{\mathrm{ref}} = \frac{2}{a}\sqrt{\frac{2E_n}{m}},
\end{equation}
if the floor potential height fluctuates by $\Delta$, the probability that a neutron in the $n$-th state is lost per unit time $t_0$ is
\begin{equation}
    \mu_n =
    \frac{t_0}{t_\mathrm{ref}}
    \left[
    \int_{U_0-E_n}^{\infty} \mathrm{d}E
    \frac{k}{\sqrt{2\pi}\Delta}
    \exp\left(-\frac{E^2}{2\Delta^2}\right)
    \right],
    \label{eq:mu_n}
\end{equation}
where $k$ is the loss coefficient. Thus, the coefficient $c_n$ in Eq.~\eqref{eq:Psi_entrance} becomes
\begin{equation}
    c'_n = c_n \exp \left( -\mu_n \tau_0 \right).
\end{equation}

We fit the theoretical wavelength distribution, including this additional loss, to the data (Fig.~\ref{fig:1D}). The fit parameters are $\theta$, $w$, $k$, and $\Delta$. Here, $w$ determines the overall normalization by setting the phase-space density of the beam. The fit excludes $\lambda>\qty{0.62}{nm}$, which is removed by the filter mirror. The best-fit results are
$\theta=\qty{-0.01(1)}{mrad}$,
$w=\qty{77.5(5)}{\micro m}$,
$k=1.4(9)$,
$\Delta=\qty{12(1)}{neV}$,
indicating that the mirror potential effectively fluctuates by about \qty{13}{\percent} relative to $U_0$ on reflection.

One can exploit the interference fringes more fully by examining the shape of the fringes in two-dimensional $\lambda$--$\phi$ space. In Fig.~\ref{fig:1D_slice}(a), we compare the wavelength distribution in the region $\left|\phi\right|<\qty{0.1}{mrad}$ between experiment and theory. We treat the angular span of the mirror $\phi_0$ as a fit parameter. Fitting yields $\phi_0^{\mathrm{fit}}=\qty{16.004(1)}{\degree}$, intermediate between $\phi_0^{\mathrm{surf}}$ and $\phi_0^{\mathrm{geom}}$. Figures~\ref{fig:1D_slice}(b) and (c) show the wavelength distributions near $\phi=\pm\qty{2}{mrad}$ using the fit result. In the negative-angle region of $\qty{-2.1}{mrad}\leq \phi < \qty{-1.9}{mrad}$, the measured neutron counts are lower than the theoretical prediction, likely because the mirror surface extends even beyond $\phi_0$ near the edge so that neutrons with $\phi<0$ (indicating motion in the $-z$ direction) lose intensity upon reflection of the mirror.

The angular span of the concave mirror was determined using three different methods: surface measurement, geometric analysis, and interference fringe fitting. The values obtained from these methods agreed to within \qty{1.3}{\percent}: $\phi_0^{\mathrm{surf}}=\qty{15.86(2)}{\degree}$, $\phi_0^{\mathrm{geom}}=\qty{16.06(2)}{\degree}$, and $\phi_0^{\mathrm{fit}}=\qty{16.003(2)}{\degree}$.
Their discrepancy exceeds each stated systematic uncertainty but is still within \qty{1.3}{\percent}.
This correspondence translates to \qty{1.9}{\percent} for the centrifugal acceleration $a$ and \qty{1.3}{\percent} for the energy scale $E_0$ because the interference pattern is determined by $E_0 R \phi_0 / v \propto E_0 \phi_0$ and the energy scale is $E_0 \propto a^{2/3}$.
Fitting of the interference fringes inside $\left|\phi\right|<\qty{0.1}{mrad}$ suggests an ideal sensitivity of \num{8e-5} to $\phi_0$ for a 5-hour measurement, or two orders of magnitude better than the present analysis. This corresponds to a potential sensitivity of \num{1e-4} for $a$ and $E_0$, which is better than the current precision of gravitationally bound states~\cite{Cronenberg_2018}. This sensitivity can be achieved if the angular span and the curvature radius of the mirror and the neutron wavelength are determined with sufficient precision and if the effects of the concave edge are accurately modeled.

\section{Conclusion and prospects\label{sec:conclusion}}
We conducted a neutron whispering gallery experiment by utilizing a pulsed cold neutron beam at J-PARC onto a precisely polished glass concave mirror and observing the bound state with a two-dimensional, time-resolving detector. We found good agreements between data and theory in the wavelength distribution by introducing an additional loss mechanism where the mirror potential height fluctuates, causing neutrons to leave the bound state. Three independent determinations of the angular span of the mirror gave
$\phi_0^{\mathrm{surf}}=\qty{15.86(2)}{\degree}$,
$\phi_0^{\mathrm{geom}}=\qty{16.06(2)}{\degree}$, and
$\phi_0^{\mathrm{fit}}=\qty{16.003(2)}{\degree}$.
They agree within \qty{1.3}{\percent}, exceeding their stated systematic uncertainties but consistent at that level. We interpret the discrepancy as arising from unmodeled deviations of the shape of the mirror edge from an ideal one. This agreement of \qty{1.3}{\percent} corresponds to verification at \qty{1.9}{\percent} for the centrifugal acceleration $a$ and \qty{1.3}{\percent} for the energy scale of the whispering gallery state $E_0$. Our analysis of the interference fringes suggests that the sensitivity could be improved to \num{1e-4} which exceeds the current sensitivity of the Earth's gravitational bound state if the angular span and the curvature radius of the mirror, and the neutron wavelength can be known more precisely and the effect of the concave edge can be correctly calculated.

For future improvements, by measuring the phase-space density of the beam independently with a high-precision collimator system, the uncertainty associated with $w$ can be eliminated. The effects of a rounded mirror edge could be more precisely calculated by simulating wave-packet evolution under a time-dependent potential boundary.
Regarding additional losses, one might minimize them by using a mirror with extremely low surface roughness or examining multiple samples with different surface conditions to identify the physical origin of the potential height fluctuations.
Furthermore, by detecting short-lived components that penetrate the mirror~\cite{Nesvizhevsky_2010_NJP}, it is possible to determine whether the additional loss is due to penetration or reflection from the mirror.
Searching for Yukawa-type corrections to Newtonian gravitaty with a range of about \qty{10}{nm} beyond the current limit~\cite{Heacock_2021,Harris_2000} requires a precision of about \num{1e-5} in $E_0$ with a high-density coating (e.g., Au or Pt)  on the mirror surface.
To achieve that, it may be necessary to measure detailed structures in the wave function such as the classical period of motion or the quantum revival period~\cite{Robinett_2004} of the neutron traveling on the mirror with a monochromatic neutron beam.

\begin{acknowledgments}
We would like to thank N.~Naganawa and Y.~Nambu of Nagoya University for fruitful discussions on equivalence principles.
This research was supported by JSPS KAKENHI Grant Numbers 21K03594 and 24K07083. The neutron experiment at the Material and Life Science Experimental Facility of the J-PARC was performed under user programs (Proposal Nos. 2021B0141, 2022A0204, and 2022B0327) and an S-type project of KEK (Proposal No. 2019S03).
\end{acknowledgments}

%

\end{document}